\documentclass[runningheads]{llncs}
\usepackage{dirtytalk}
\usepackage{graphicx}
\usepackage{float}
\usepackage{cite}
\usepackage[colorinlistoftodos,prependcaption,textsize=tiny]{todonotes}
\begin{document}
\title{A Document-based Knowledge Discovery with Microservices Architecture}
%
%
\author{
Habtom Kahsay Gidey \and
Mario Kesseler \and
Patrick Stangl  \and
Peter Hillmann  \and
Andreas Karcher
}
\authorrunning{H.K Gidey et al.}
%
\institute{
Universität der Bundeswehr München, Germany \\
\email{\{habtom.gidey, mario.kesseler, patrick.stangl, peter.hillmann, andreas.karcher\}@unibw.de}
}
\maketitle              
\begin{abstract}
The first step towards digitalization within organizations lies in digitization - the conversion of analog data into digitally stored data. This basic step is the prerequisite for all following activities like the digitalization of processes or the servitization of products or offerings. However, digitization itself often leads to \say{data-rich} but \say{knowledge-poor} material. Knowledge discovery and knowledge extraction as approaches try to increase the usefulness of digitized data. 

In this paper, we point out the key challenges in the context of knowledge discovery and present an approach to addressing these using a microservices architecture. 
Our solution led to a conceptual design focusing on keyword extraction, similarity calculation of documents, database queries in natural language, and programming language independent provision of the extracted information.
In addition, the conceptual design provides referential design guidelines for integrating processes and applications for semi-automatic learning, editing, and visualization of ontologies.
The concept also uses a microservices architecture to address non-functional requirements, such as scalability and resilience.
The evaluation of the specified requirements is performed using a demonstrator that implements the concept. 
Furthermore, this modern approach is used in the German patent office in an extended version.
\end{abstract}
\section{Introduction}
Digitization coupled with fast-paced advances in various areas of computing has resulted in an unprecedented volume of data. 
Every application produces volumes of data for which usable information must be searched and analyzed. 
This data growth, in return, challenges existing knowledge systems in knowledge-based organizations. 
Taking the intellectual property (IP) institutions as an example, the European Patent Office has experienced a boom in technical patent applications since the last economic crisis in 2009.
Patent applications have increased by more than 34 \%~\cite{StatEPO2020}.
Success in digitization has also intensified the increase in the patent examination workload by changing how IP applications are submitted and processed. 
While almost 100 \% of all patent applications were filed in writing in 2004, nearly 90 \% of submissions are now made digitally~\cite{Deuc2019}. 
Processing within the office is fully digital.
The figures are likely similar for many other property rights.
The growing number of applications also increases the workload for patent offices worldwide. 
In particular, the search, retrieval, and examination workload, which takes up the largest share of time in granting a patent, increases with each patent application. 
For example, some patent offices completed more than 40,000 examination procedures in 2019, which is a significant increase in workload in patent examinations~\cite{Deu2020}. 
This contrasts with more than 67,000 new patent applications in the same year.  
This time expenditure signifies the high level of knowledge that patent examiners must maintain to carry out their daily work.  
It starts with the classification of patent applications according to international classification specifications, particularly the search for similar patents. 
Then, a patent examiner must complete searches and examinations not to grant the IP right erroneously, requiring corresponding domain knowledge for the examiner. 
Consequently, the training period of a patent examiner is five years before he can perform patent examinations entirely independently.
Previous attempts to raise the number of examinations processed have always been to increase the number of patent examiners.

However, the goal must be to use knowledge systems to support the knowledge worker and change the \say{data-rich}, but \say{knowledge-poor} scenario by reducing the processing times for classification, search, and examination, and the training time for new patent examiners. 
Intellectual property institutions process digitized unstructured documents instead of structured information. 

As a result, this paper presents a contribution that addresses the knowledge systems challenges in document-based knowledge discovery (KD) with a highly flexible microservices architecture. 
The paper is organized as follows. 
The background and related work is presented in Section II. Section III describes the conceptual model for KD. Section IV presents the evaluation of the model, and the final section presents the summary and an outlook.
\section{Research Context}\label{sec:ResearchContext}
KD is a topic of broad scientific interest in information systems research.
Automated processing of unstructured data classification, retrieval, and testing is of particular interest for this work.
In this context, the knowledge system for KD is set around the processes of classification, search, and examination of patent applications. 
The classification of a patent application serves the correct assignment of a patent submitted to the responsible examining office. 
For this purpose, the International Patent Classification (IPC) scheme is applied, which provides uniform hierarchical classes and specific sub-classes~\cite{Wor20b}. 

First, patent applications are classified roughly into the relevant classes during the classification process. 
Next, based on the classification, the applications are submitted in a round-robin procedure to the presumably competent preliminary examiner. 
Then, the latter either confirms the classification carried out, refines it according to sub-classes, or determines that the assigned class is incorrect. 
In case of an error, the final classification is determined by the auditors of other possibly competent auditing bodies.      

The classification of the patent applications is based on the intellectual registration of the contents of the respective patent specification. 
In particular, the claims, descriptions, or attachments are of specific interest.  
The focus of the intellectual content to be examined varies from one examination area to another. 
For example, in the case of applications in chemistry, the representations of chemical compounds are decisive, whereas, in electrical engineering, the claims for classification are more important.

A search process always precedes the patent examination process, which looks for similarities among patent applications. 
In general, this process follows a sequence characterized by the intellectual acquisition of the contents of the patent application and the search for already filed applications with similar contents.
These are intelligent comprehension of the contents of the new patent application, compilation keywords of the technical concepts that characterize the described patent, search of referenced documents or documents that in turn reference this application, and then search for the assigned keywords in the documents of the corresponding IPC class.

Similar to the classification, the contents essential for the search differ depending on the examination field. 
When searching in the respective IPC classes, up to 2000 documents may have to be searched for similar contents and concepts. 
The concepts can be realized by technical drawings or defined descriptively by terms and relations to other terms. 
The search for similar concepts also explains the high time expenditure of a patent examination.
\subsection{Example Scenario}
The following scenario describes the vision of the new examination process of a patent application as it appears after a potential deployment of an exemplary KD system:
\textit{
Julia is one of 700 examiners and 2000 other employees at a hypothetical patent office. 
She works at different office locations, including from home. 
Currently, she is a trainee investigator and has to process the new applications assigned independently as part of her training. 
She has just received a patent application from a company on cognitive systems. 
Since all new applications are automatically classified and assigned keywords when they arrive at the office, the system notifies her immediately after submitting the patent application.
Opening the patent specification document, she gets live support with various keywords that capture the core of the patent application. 
Julia then reads the relevant passages of the patent specification and determines the classification recommendations are valid.
She now has two different ways to start searching for similar patent applications. 
In the first case, she receives a list of all other applications sorted by relevance to which comparable keywords have been assigned.  
In the second approach, she actively searches for comparable content. 
To minimize the training effort, she can ask the question in a natural language. 
Julia thus writes into the search field: `Show me all applications with the keyword cognitive systems.'  
She receives a list of all applications containing the keywords, sorted by relevance.
In both cases, she has an up-to-date and limited list of documents based on which she can make an intellectual comparison of the keywords without having to click through several hundred patent applications. 
Since Julia is still a trainee, she needs to understand the interrelationships of patents in cognitive systems before comparing individual patent applications. 
For this purpose, she looks at an ontology provided by the system, which represents the concepts and relationships between cognitive systems and other related topics - such as cognitive models and cognitive architectures. 
After comparing submitted patent applications with the new patent application, Julia decides on the novelty of the patent application on a new cognitive architecture.}
\subsection{Research Questions}
We posed the following research questions (RQs) to conceptualize and evaluate the document-based KD solution. 
\noindent\textit{\textbf{RQ1}: Considering the increasing importance of IP rights, what are the challenges for knowledge workers in existing workflows of a patent application and examination?}
\noindent\textit{\textbf{RQ2}: What are the main aspects of knowledge systems that address practical KD requirements in processing and examining patent applications?}
\noindent\textit{\textbf{RQ3}: What are the ways to realize architecturally significant requirements of a future-proof document-based KD system in patent classification and examination?}
\section{Background and Related Works}
Architectural approaches and design decisions make significant contributions toward making software systems scalable, resilient, and future-proof~\cite{furrer2019future,Jansen2005,gidey2017grounded}. 
The microservices architecture (MSA) is, for instance, an architectural pattern that has demonstrated value in addressing the challenges caused by the increased need for rapid digitalization and servitization in data-rich domains~\cite{kohtamaki2020relationship,vandermerwe1988servitization}.
The MSA separates application services based on business capabilities or a domain's functional requirements~\cite{richardson2018microservices}.
Services are then restricted on domain context and size~\cite{dragoni2017microservices}. 
They are also deployed, managed, and scaled independently of each other. 
Similarly, services communicate with each other independently with messaging protocols such as HTTP/REST.~\cite{garriga2017towards, richardson2018microservices}.
Due to the strongly decoupled micro-sized services, MSA's software components are easy to maintain or even replace.
As a result, MSA has also been a preferred path for architecture-driven software modernization~\cite{knoche2018using}. 
MSA further addresses the architectural challenges of KD by componentizing the extensive functional domains such as natural language processing, keyword extraction for text mining, and ontology management into services~\cite{dragoni2017microservices}. 
As a process of useful knowledge extraction from unstructured documents, document-based KD has distinct technical requirements that differ from other types of KD.~\cite{ahonen1999knowledge}.
The key differences lie in the requirements for which the need is `information retrieval' vs. `text mining.' 
According to Ben-Dov et al., this topic corresponds to the sub-area `information retrieval,' which is concerned with finding new information across individual data records ~\cite{ben2005text}. 
Hotho et al.~\cite{hotho2005brief} also show the lack of clarity in the definition of the term itself, which ranges from information extraction to an all-encompassing `knowledge discovery' process.

Besides, MSA has grown as an architecture of choice for diverse applications in software development practices~\cite{di2017research}. 
Although literature presents several MSA implementations for a growing number of software solutions, very few select exist that address architectural challenges in KD services. 
Singh et al.~\cite{singh2020mubigmsa}, for example, have implemented a reference application based on a microservice-based model, which they proposed for Big Data KD. 
Vekaria et al.~\cite{vekaria2020recommender} also have presented a chatbot-based recommender system for Science gateways to support KD with augmentable modules as microservices. 
However, those few examples have no focus on the challenges of document-based KD and significantly differ from our solution of document-based KD with an MSA.
\section{Conceptual Approach}\label{sec:ConceptualApproach}
In the following, we conceptualize the essential aspects of KD requirements with microservices architecture to address knowledge-intensive document processing challenges.
The conceptual structure determines the individual microservices' specification, the necessary data model, and each service's data persistence. 
Subsequently, the communication between the individual services is also defined. 
Furthermore, individual microservice size and functionality are also briefly described.
We also describe the design decisions made throughout the conceptualization, alternative design rationales considered, advantages and disadvantages of each possible design choice, and the selected design justifications.
\subsection{The Microservices Specification}
We specify the necessary microservices into two parts. 
The first part specifies the microservices related to the domain, KD requirements. 
The second part specifies the services significant for the infrastructure of the microservices architecture.
\subsubsection{Domain Related Microservices:} 
\begin{figure}[htbp]
    \begin{minipage}[t]{0.425\linewidth}\centering
        \includegraphics[scale=0.3]{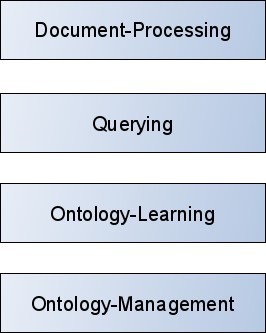}
        \caption{Subdomain structure}\label{fig:subdomain}
    \end{minipage} 
    \begin{minipage}[t]{0.48\linewidth}\centering
        \includegraphics[scale=0.3]{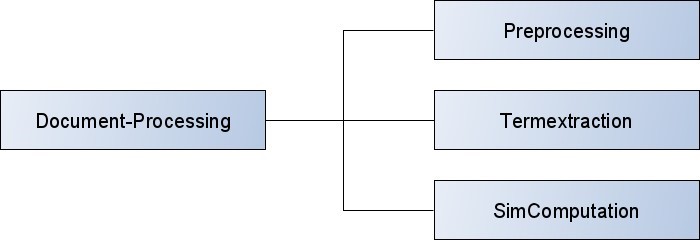}
        \caption{Document processing structure}\label{fig:docProcess}
    \end{minipage}
\end{figure}

In Fig.~\ref{fig:subdomain}, following Domain-Driven Design~\cite{evans2004domain}, we identify domain-related microservices from the four subdomains. 
The document processing subdomain contains the functionalities for determining keywords and calculating the similarity of documents. 
The Querying domain provides all query mechanisms for standard queries for keywords and queries via Natural Language Interfaces. 
The domain Ontology-Learning contains the missing layers for automatically extracting ontologies from unstructured text.
Since term extraction has already been done in the Document Processing domain, all tasks are built on top of it after the Ontology Learning Layer Cake.
The last domain to be mentioned is Ontology Management, which allows the editing and visualization of automatically generated ontologies.
The subdomains form self-contained units, which have the following advantages:
\begin{enumerate}
\item Self-contained, independent data models,
\item Independent scalability of each subdomain, 
\item Internal subdomain changes do not affect the entire system.
\end{enumerate}

Independence of the individual domains at runtime, for example, queries via Querying, can be performed without functioning document processing. 
However, independence does not mean that the system can work in a meaningful way without processed documents.
A filling with documents from which keywords and ontologies are extracted is a prerequisite for the system's usability. Following the top-down approach, these coarse sub-domains can now be more precisely subdivided into `sub-sub domains.' 
Thus, a structure can be determined for the processing of the documents, as shown in Fig.~\ref{fig:docProcess}. 
Here, the service `Preprocessing' takes over converting the specified file formats into pure, machine-readable text. 
The service `Termextraction' extracts the keywords from the text, and the service `SimComputation' computes the similarities between the newly added documents and the documents already existing in the system. 
Based on the controller and reporter pattern, the parent service Document-Processing handles the processing control and maintains the documents' status.

An alternative approach for the similarity calculations would be to determine the similar documents only at the time of a potential request to the system - i.e., only when similar documents are to be output for a given document. 
However, this has a disadvantage: the user expects a prompt, timely response to a request. 
Similarity computations are performed only at a point in time, resulting in higher requirements on the performance of the service, which has to process several requests simultaneously and perform the computations.
An improvement of the situation at the time of a request by a user would occur if the service to which the request goes does not compute the calculation at the time of the request but at the time of integration of a new document.
However, the approach chosen here also has the advantage over the approach in the service in which the queries are processed that only database operations are running, and no additional computations have to be performed. 
This ensures the highest possible performance for queries by users.

\begin{figure}[htbp]
    \begin{minipage}[t]{0.45\linewidth}\centering
        \includegraphics[scale=0.25]{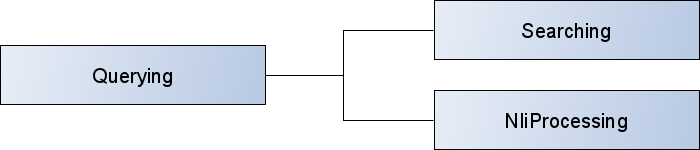}
        \caption{Querying subdomain}\label{fig:querying}
    \end{minipage} 
    \begin{minipage}[t]{0.45\linewidth}\centering
        \includegraphics[scale=0.25]{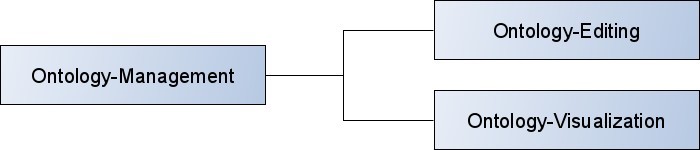}
        \caption{Ontology management}\label{fig:om}
    \end{minipage} 
\end{figure}

\begin{figure}[htbp]
    \begin{minipage}[t]{0.44\linewidth}\centering
        \includegraphics[scale=0.25]{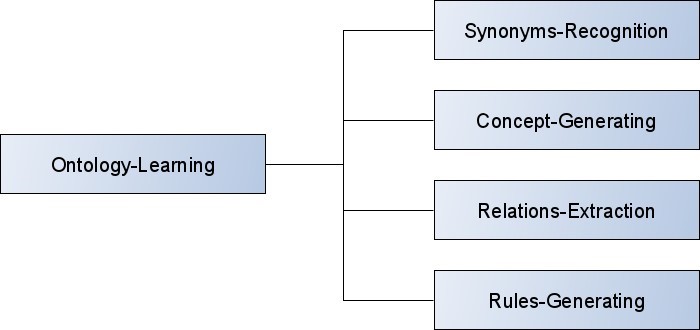}
        \caption{Ontology-Learning subdomain}\label{fig:ol}
    \end{minipage}
    \begin{minipage}[t]{0.54\linewidth}\centering
        \includegraphics[scale=0.17]{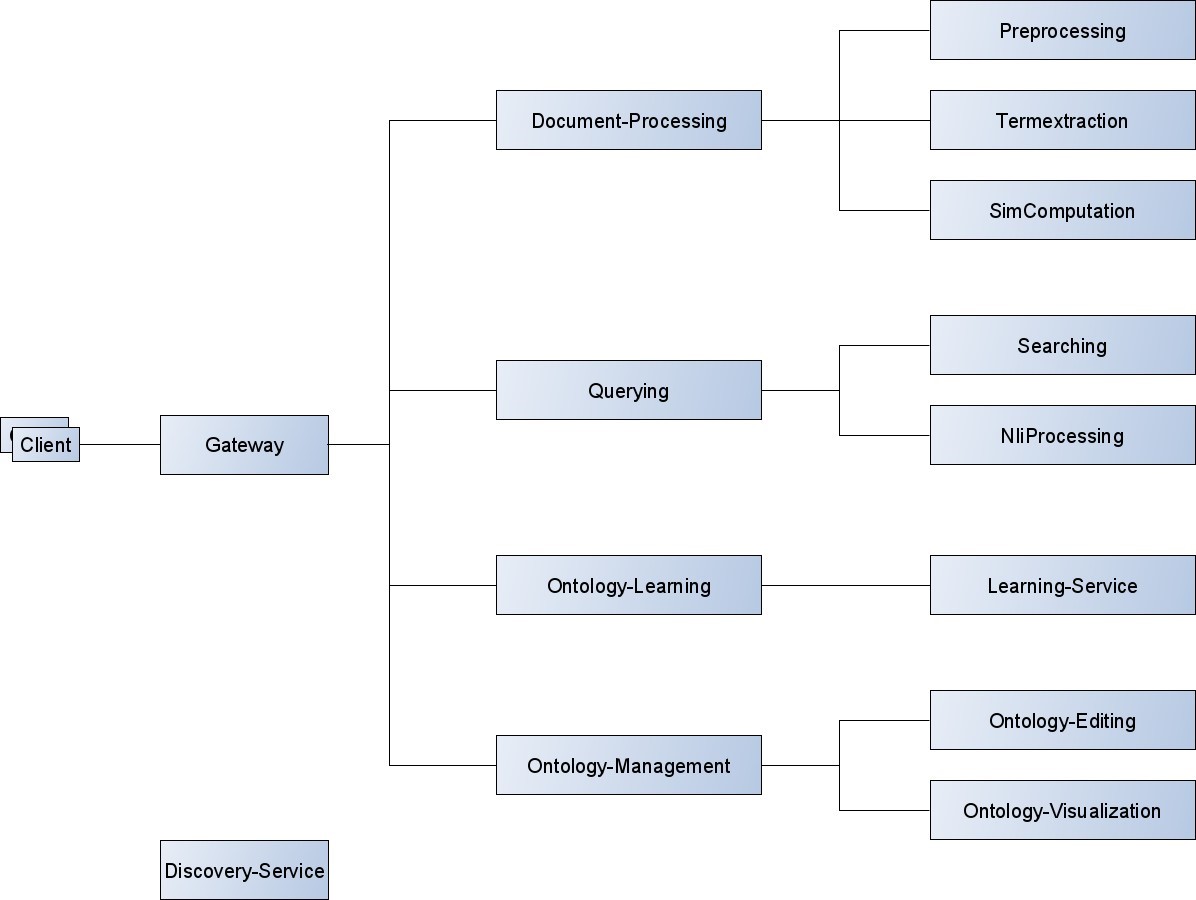}
        \caption{System structure with components}\label{fig:concept}
    \end{minipage}
\end{figure}

For the subdomain `Querying,' a microservices structure results as shown in Fig.~\ref{fig:querying}.
Here, the individual subordinate services provide different possibilities for searching. 
The service `Searching' provides web services for standard queries for similar documents or keywords. 
The service NliProcessing, on the other hand, processes queries formulated in natural language. 
The parent service `Querying' bundles the queries, can be used to provide caching for searches and search results, and is furthermore available for aggregating search results from the different search possibilities. 
Caching within the parent service allows very fast response times. 
At the same time, the splitting into individual subordinate services ensures that in case of a failure of a sub-service not all queries come to nothing.
Taking the ontology learning layer cake as a basis for the ontology learning subdomain and defining the subordinate microservices according to the presented layers results in the exemplary service structure shown in Fig.~\ref{fig:ol}.
After determining the synonyms for the extracted keywords in the service `Synonym- Recognition,' the concepts are determined in the service `Concept-Generating,' and the relations between the individual concepts are extracted in the following service `Relations-Extraction.'
The rules or axioms, which can be derived automatically from the previous information, are created in the service `Rules-Generating.' 
As can be seen, the service for keywords extraction is omitted here since this has already been performed in the document processing context.
Since the existing and outdated solutions of ontology learning perform the whole task of ontology learning as a black box in an application, the problem arises that an ideal-type separation into the layers of the ontology learning layer cake seems impractical. 

The structuring into a superordinate and a subordinate service allows the integration of a single ontology learning framework into the learning service or an ideal-typical implementation where the ontology learning service takes over the control of the workflow. 
For the last subdomain, ontology management, a microservices structure results as shown in Fig.~\ref{fig:om}. 
The automatically generated ontologies additionally require monitoring and editing capabilities. 
Editing capabilities are provided by the service `Ontology-Editing,' whereas the service `Ontology-Visualisation' provides visualization possibilities for the presentation of the ontologies.
The higher-level service `Ontology-Managment' is primarily used for forwarding the queries and the learned ontologies. 
\subsubsection{Infrastructure Related Microservices:}
Loose coupling between the individual microservices is a fundamental advantage of the microservices architecture. 
At the same time, it leads to the need to integrate services for localizing the individual microservices. 
For the localization of all microservices instances, a registration and discovery service is necessary, which is continuously informed about available instances of services and returns an instance of the desired service to the caller upon request. 
A gateway service must also be integrated to hide non-public interfaces from external clients. 
Additionally, other infrastructure services can be integrated, such as authentication. 
After the integration of the newly added infrastructure components, the overall system corresponds to Fig.~\ref{fig:concept}. 
In addition to the functional and infrastructural services listed here, further services for persistence and asynchronous communication are necessary. 
These will be integrated successively in the following chapters.
\subsection{Data Model}
The data model of the conceptual approach is divided into an internal and an external data model. 
The internal data model is used for data processing within the individual microservices and data transfer between the microservices. 
The external data model is intended to be used for communication with the clients who access the rest of the web services provided.
\subsubsection{Internal Data Model:}
The services data model results altogether from the required input of the service and its output, which can represent the input of another service or the information requested by the user.
For this task, a distinction must be made between the subdomains of document processing or querying and the subdomains focusing on learning and managing ontologies.
For the subdomains Document-Processing and Querying, an overall data model is composed of the data models of the subordinate services. 

In addition, it requires the ability to uniquely identify a file and the information derived from it across the individual microservices. 
For this purpose, an ID must be assigned while uploading a document, which is available in the data model of all further microservices. 
The algorithm used must also be provided to the services' output to meet different keyword extraction and similarity calculation methods requirements.
For the data model of the ontology-related services, it should be noted that the data input for ontology management and ontology learning services is different. The data model of the Ontology-Management service is the ontology itself. 
Since this service provides methods or applications for manual editing of ontologies, the data model of the input also corresponds to the data model of the output. 
In both cases, editing and visualization ontologies are involved.

This restriction does not apply to the ontology learning service. 
Here, two factors have to be taken into account: first, the data model of a newly processed document from which further information for new or existing ontologies shall be extracted by processing in the ontology learning process, which corresponds to the data model of the term extraction service and second, the data model of the ontology that is either already exists due to automatic ontology extraction of previous runs or is provided to the ontology learning microservice after a manual modification via the ontology editing microservice.

Thus, the picture for the ontology-learning microservice is that both a data model for reproducing a new document and a data model for reproducing ontologies are necessary. 
For this purpose, the Web Ontology Language is chosen. 
This decision is equally viable if a microservices structure is envisaged, following the Ontology Learning Layer Cake layers. 
OWL is also used as a data model for the ontology management subdomain since existing applications support this data model either directly or via plugins.
\subsubsection{External Data Model:}
A different approach has to be chosen for the data model used for the communication with the clients. 
Decisive here is the web services which are made available to the client. 
In this concept instance, a distinction has to be made between document processing, querying, and ontology management domains. 
The domain Ontology-Learning only processes data internally and has no external interfaces. The domain Ontology-Management allows the processing and visualization of ontologies. 
Since only existing applications like WebProtege or WebVOWL are linked here, reference shall be made to their possibilities to download data. 
In Querying, a distinction has to be made between Searching and NliProcessing. 
While the Searching Microservice only allows predefined queries defined in requirements, the NliProcessing Microservice allows free text input.
Since no exact data structure of the response is known under these conditions, a generic data model must be implemented. 
\subsubsection{Persistence:}
\begin{figure}[htbp]
    \begin{minipage}[t]{0.45\linewidth}\centering
        \includegraphics[scale=0.22]{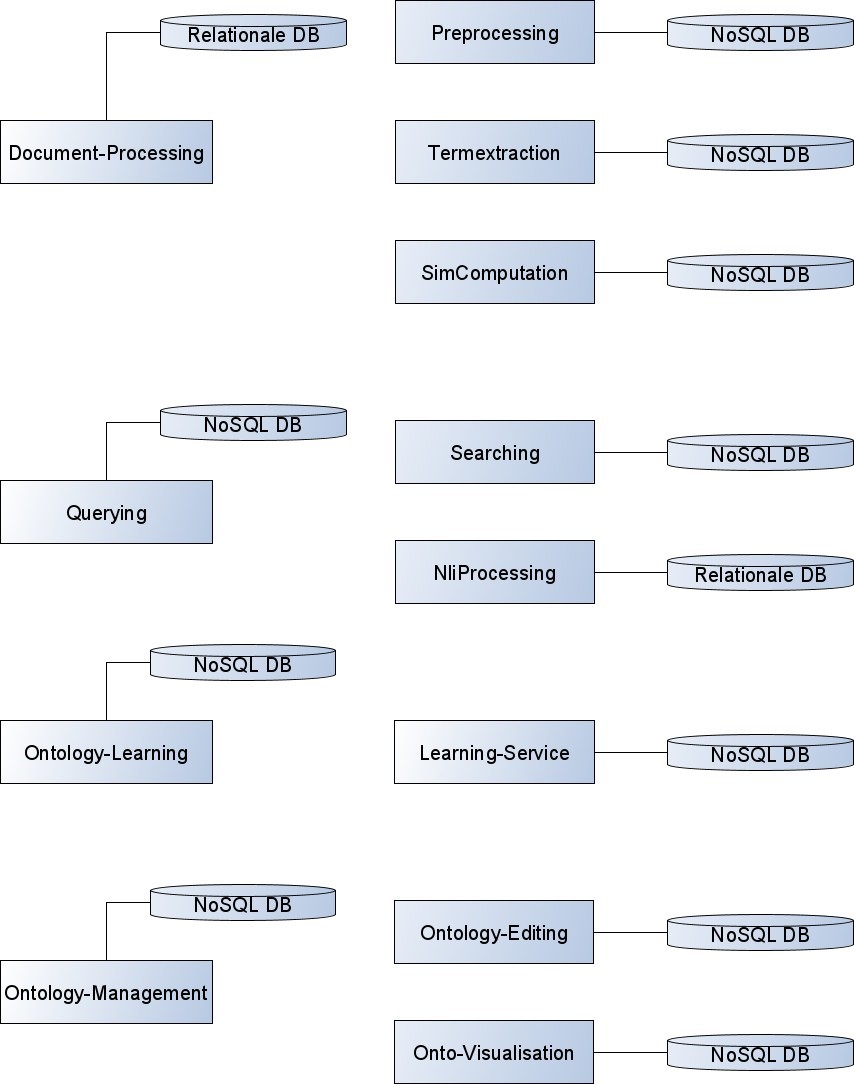}
        \caption{Databases for the microservices}\label{fig:db}
    \end{minipage} 
    \begin{minipage}[t]{0.48\linewidth}\centering
        \includegraphics[scale=0.2]{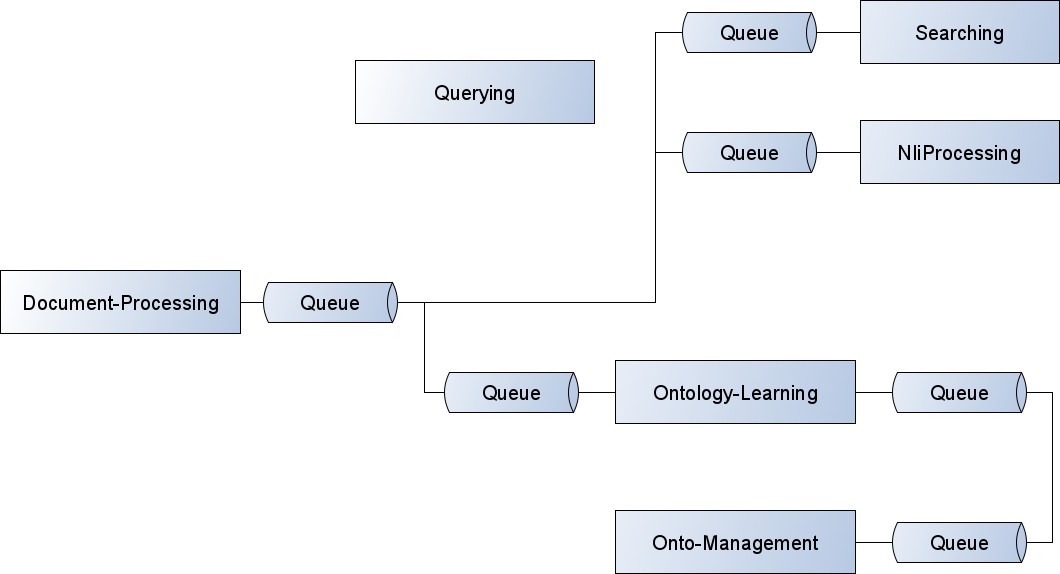}
        \caption{Cross-domain communication}\label{fig:crossDomainComm}
    \end{minipage}
\end{figure}

To conclude the chapter on data models, the persistence of the data in the individual microservices shall be considered. 
Here, a basic decision between relational and NoSQL databases has to be made for each microservice. 
The question of the type of database used depends primarily on the structure of the data to be stored. 
The simpler the data structure, the better the data can be mapped into an SQL database without complex joins. 
The more complex or flexible the data structures are, the easier it is to store data in NoSQL databases. 
This is also true for large amounts of text or when the data structure to be stored changes. NoSQL databases have supported both aspects since their creation. 
The full support of reactive programming by NoSQL databases is another advantage, especially in areas where performance plays an important role. 
The databases shown in Fig.~\ref{fig:db} are chosen for the persistence layer for this concept considering the reasons above. 
The schemas of the individual databases are based on the classes of the internal data model. 
Persistence in caching in the Querying Microservice relies on a NoSQL database since different results with different structures have to be kept. 
In the NliProcessing microservice, on the other hand, the use of an SQL database is necessary. 
Here, however, it is additionally important to adapt the database schema to the possibilities of the service to convert text into SQL commands since not all SQL constructs are supported. 
In all microservices of the ontology management sub-domain, the ontologies in OWL format are also kept in a no-SQL database.
\subsection{Communication}\label{Communication}
After conceptualizing the data models, we defined and specified the communication between the individual services on three levels. 
These are, first, the communication between client and public interfaces, 
second, the cross-subdomain communication within the microservices and, finally, the data transfer within the domain-specific microservices structures. 
The specification is limited to functional communication. 
Moreover, communication to the infrastructure services is exclusively synchronous and will not be considered.
\subsubsection{External communication:}
A client accesses the public interfaces via the web services provided by the system. 
The services mentioned are offered in the form of URLs and accept data synchronously or deliver information synchronously. 
These are transmitted according to REST-compliant standards, i.e., marshaling to or unmarshalling from JSON takes place. 
Only standard HTTP methods such as POST and GET are used for communication. 
This applies to all public interfaces, except for the interfaces provided in the ontology management domain, since there are no web service-compatible applications that could be integrated. 
The interfaces provided have to be used to integrate an existing ontology management software like WebProtégé. 
This also applies to the visualization component to be integrated. 
The decision in favor of synchronous communication between client and system is based on the expectation of a timely and high-performance response, even though asynchronous communication is preferable in microservices architectures.
\subsubsection{Cross-domain technical communication:}
Communication within the system across the individual subdomains is asynchronous since this type of data transfer creates additional decoupling. 
This requires an additional messaging service in the system that provides the necessary queues. 
The standard JSON is again used as the data transfer format, and the FIFO principle is used to process the messages in the queues. 
The entire cross-domain communication then corresponds to Fig.~\ref{fig:crossDomainComm}. 
Since the further processing or provision of the data within the subdomains Querying and Ontology Learning depends on the preprocessing within Document Processing, the data is forwarded asynchronously from Document Processing to the services mentioned earlier. 
It should be noted that querying the data is not forwarded through the higher-level Querying microservice to the lower-level Searching and NliProcessing microservices but directly to the latter two.
This is done with regard to the performance and error resilience of the querying services. 
This approach ensures that the querying service is only burdened by the requests from the outside and does not have to process the forwarding of the newly processed documents additionally. 
Second, this means only two instead of three queues are needed in the messaging service, and the network load is reduced by a third.
Third, it allows new documents to be integrated into the two lower-level microservices even if the querying service is down or overloaded.

This approach is not target-oriented when providing data for the Ontology-Learning service. 
Since this microservice does not provide any public interfaces, allowing for future changes regarding the pragmatic approach described above or a split along with the layers of the Ontology Learning Layer Cake, the asynchronous delivery of the data to the parent Ontology-Learning Microservice is to be preferred. 
Furthermore, for the cases considered so far, the response of successful processing to document processing is also asynchronous. 
The last thing to specify is the communication between the ontology-learning and ontology-management microservices. 
The ontology management applications do not need direct input from the document-processing microservice but only the ontologies elaborated in the ontology-learning subdomain.  
At the same time, the ontology management, in turn, has to provide the ontologies revised manually to the ontology learning microservice. 
Thus, data exchange has to take place in both directions, whereby asynchronous communication is also preferred in each case.
\subsubsection{Communication in document processing:}
In the Document-Processing subdomain, internal communication with the subordinate microservices occurs in both asynchronous and synchronous forms. 
The document processing microservice accepts the documents uploaded by the client synchronously and forwards the result of the entire processing to the querying and ontology learning subdomains asynchronously. 
The processing of the documents takes place in individual steps, which are broken down into subordinate microservices. 
The communication can be differentiated into the normal processing flow and the data retrieval in the event of an error. 
The document processing service provides the individual subordinate microservices with the required input data asynchronously and receives the result asynchronously at the end of the respective processing step. 
In the event of an error, the subordinate microservices provide synchronous interfaces to return the processing result of a document that has already been processed. 
This allows the document processing service to fall back on the result of the previous processing step if errors occur and feed it back into the normal workflow. 
\subsubsection{Communication in Querying:}
The communication in the Querying subdomain is a purely synchronous data transfer based on the Querying microservice as an intermediary to the respective subordinate microservices. 
In addition to the publicly provided interfaces, the services in this domain do not include any other interfaces.
\subsubsection{Communication in Ontology-Learning:}
Communication in the ontology-learning subdomain is structured asynchronously.
The reason for this is the potentially long runtime of the generation of ontologies and the individual steps involved. 
It should also be noted that an asynchronous workflow can also be set up in parallel to the procedure in document processing when using individual microservices per task of the Ontology Learning Layer Cake.
\subsubsection{Communication in Ontology Management:} 
Data transfer in ontology management is also asynchronous.
The ontologies generated in the Ontology Learning domain and edited via the Ontology Editing service are provided asynchronously to other services. 

Ontologies provided by the Ontology-Learning service are forwarded asynchronously to both subordinate microservices via the Ontology Management microservice.
Ontologies that the Ontology-Editing service has edited are in turn provided asynchronously to Ontology-Learning as well as to the Onto-Visualization service via the Ontology Management service. 
In addition to this asynchronous communication, synchronous public interfaces forward the user to the web applications for editing or visualization. 

\begin{figure}[htbp]
    \centering
    \includegraphics[scale=0.3]{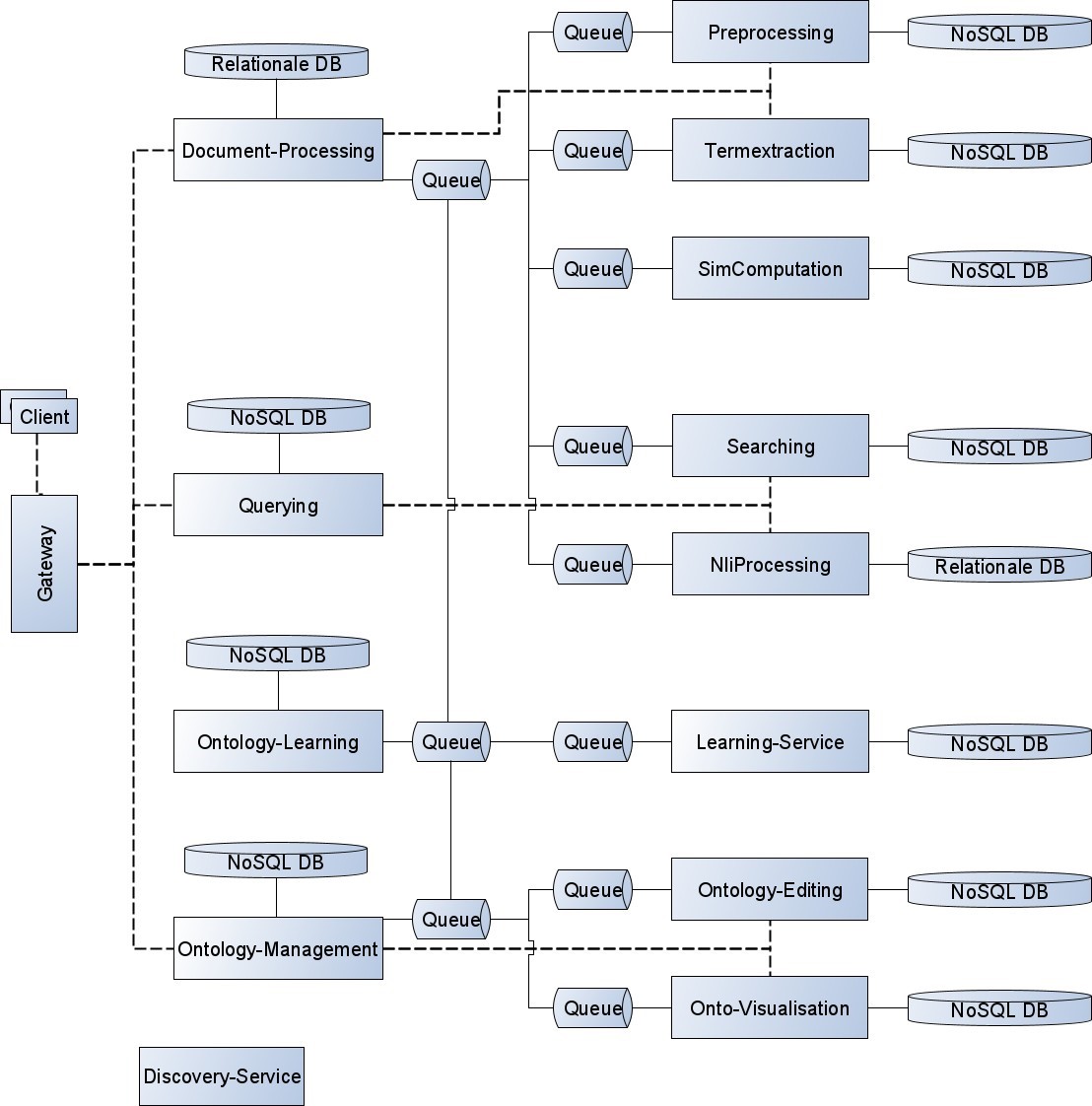}
    \caption{Overview of the system structure}
    \label{fig:sys}
\end{figure}

After integrating communication and persistence into the system, the structure shown in Fig.~\ref{fig:sys} results for the entire system. 
Solid lines with queues between the individual microservices correspond to asynchronous communication and dotted lines to synchronous communication. 
The synchronous communication of all microservices to the discovery service is hidden in the diagram.
\section{Evaluation and Assessment}
\subsection{Implementation}
The conceptual model discussed in the previous section is implemented in separate domain-relevant components where microservices are also divided into technical and infrastructure-related services. 
Besides, all the microservices described are implemented in Java using the Spring Boot Framework and Docker. 
Some parts are fully implemented, such as Document-Processing and Querying microservices. 
However, the Ontology-Management service is partially implemented using existing ontologies. 
The implementation is set up as a multi-project build containing a separate Gradle project for each required microservice. 
\subsection{Evaluation}
We evaluate our concept according to our scenario in Section II, based on typical processes at a patent office.
It includes using a microservices architecture, RESTful APIs, existing libraries, frameworks, and services already in use, with an example scenario of a patent office.  

For the testing, we used the Postman tool and cURL, which allows for sending whole collections of requests.
For instance, the requirement, which entails the application to process PDF format documents, is assessed as the first step in the success of uploading documents via the provided REST API.  
In this case, forty document packages related to patents and science were uploaded to the system, and the processing was monitored. 
The success of the upload was traced in the database successfully.

\begin{figure}[htbp]
    \centering
    \includegraphics[scale=0.28]{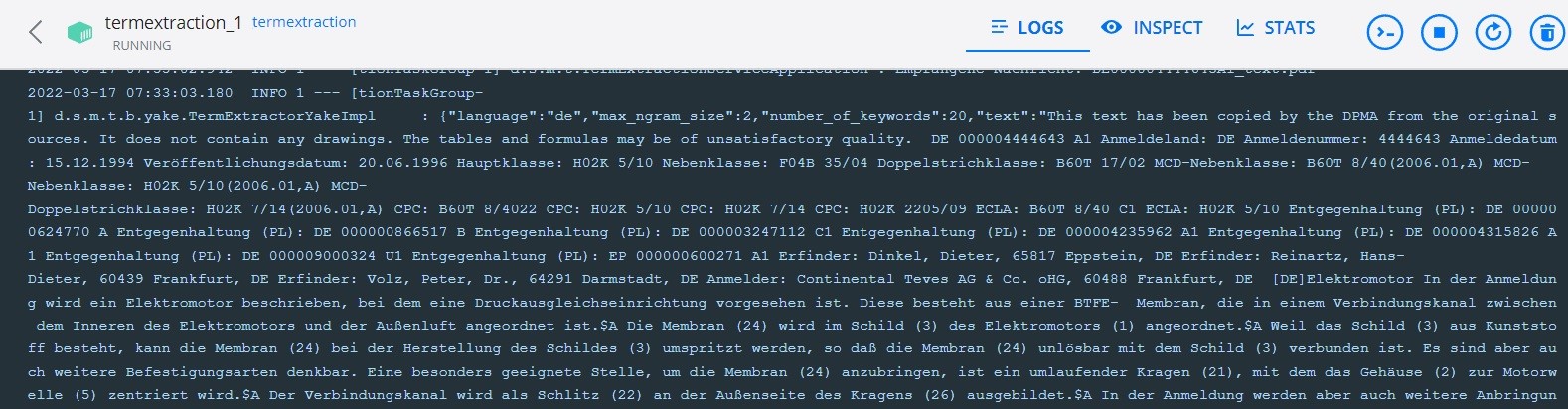}
    \caption{Extract from collection `Extractions' with extracted text.}
    \label{fig:eval63}
\end{figure}

A second requirement, which entails automatic text extraction from the available documents, evaluates the preprocessing microservice. 
Proof of successful processing can also be observed on the content of the microservice's database. 
Fig.~\ref{fig:eval63} shows a section of the `Extractions' collection after successful text extraction.
Other requirements, such as the automatic extraction of keywords from documents and presenting a requested list of keywords for each document, can be evaluated together. 

\begin{figure}[H]
    \centering
    \includegraphics[scale=0.22]{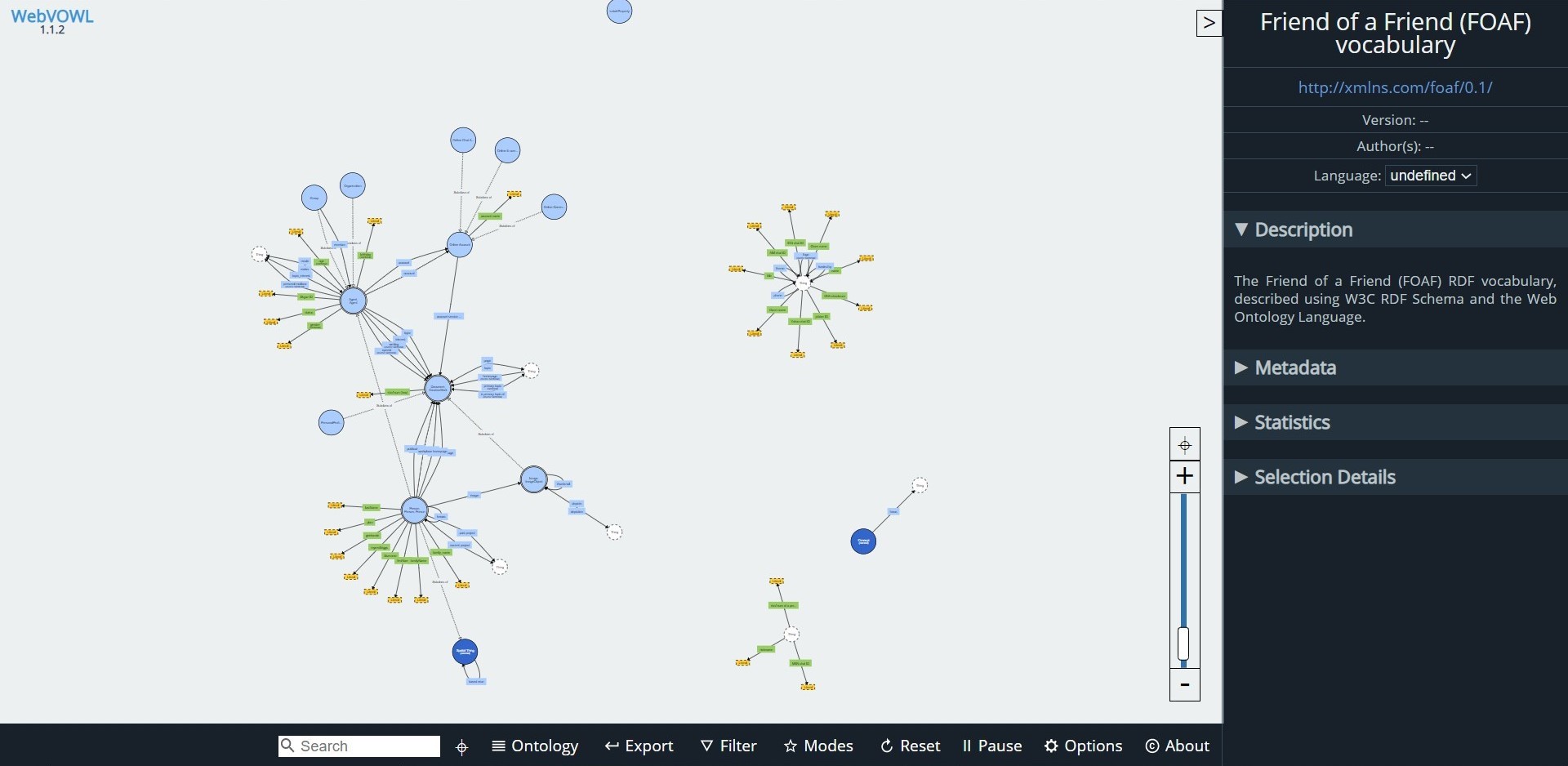}
    \caption{Redirection to WebVowl}
    \label{fig:struct}
\end{figure}

Furthermore, the evaluation of the requirement that the application must support integrating tools for visualizing ontologies is conducted assuming the ontology in OWL format exists. 
The visualization component WebVowl is called with an existing ontology in OWL format. 
Fig.~\ref{fig:struct} shows this component after calling the web service `ontomanagement/getVisualisation'.
\section{Conclusion}
KD is an integral part of knowledge-intensive organizations and their processes servitization. 
To this end, in this paper, we have presented a conceptual model and an evaluation of document-based KD with an example scenario at an intellectual property institution. 
The conceptual approach mainly focused on using microservices architecture to model and implement four main domain-relevant microservices: document processing, querying, ontology learning, and ontology management services. 
Due to its extensibility, microservices architecture is an ideal basis for creating knowledge-based applications. 
Workflows for processing documents can be easily implemented and almost completely separated from the provision of the resulting information.
As a result, keyword extraction, similarity determination, and provision of information based on a RESTful API were successfully implemented for the document-based KD.
Then, examples of key requirements were demonstrated on how implemented services were assessed and examined. 
The benefits of the KD implemented are identifying keywords from knowledge-intensive documents, supporting the recognition of similarities among them, and generating and retaining essential knowledge of the documents. 

As an outlook, future work can extend this concept from several points of view. 
An apparent investigation is the creation of a microservice-based concept of the `Ontology Learning Layer Cake.' 
Furthermore, the general consideration of NLP-based problems is interesting. 
The question to be answered here is how small-scale NLP tasks can be decomposed into individual microservices to act as part of different problem-specific NLP tasks. 
Further research based on this work could, in turn, aim to develop MSA-based NLP analysis frameworks. 
%
%
%
\bibliographystyle{splncs}

\bibliography{references}
\end{document}